\documentstyle[12pt,procsla_ismd,epsf]{article}
\catcode`\@=11
\long\def\@makefntext#1{
\protect\noindent \hbox to 3.2pt {\hskip-.9pt
$^{{\ninerm\@thefnmark}}$\hfil}#1\hfill}		%CAN BE USED

\def\@makefnmark{\hbox to 0pt{$^{\@thefnmark}$\hss}}  %ORIGINAL

\def\ps@myheadings{\let\@mkboth\@gobbletwo
\def\@oddhead{\hbox{}
\rightmark\hfil\ninerm\thepage}
\def\@oddfoot{}\def\@evenhead{\ninerm\thepage\hfil
\leftmark\hbox{}}\def\@evenfoot{}
\def\sectionmark##1{}\def\subsectionmark##1{}}
\def\bet{\begin{equation}}
\def\eet{\end{equation}}
\def\bc{\begin{center}}
\def\ec{\end{center}}
\def\ov{\over\displaystyle\strut}
\def\dst{\displaystyle\strut}

\def\t0{\tau_0}

\def\ben{\begin{eqnarray}}
\def\enn{\end{eqnarray}}
\def\l({\left(}
\def\r){\right)}

\def\t{{\tau}}
\def\o{{out}}
\def\s{{side}}
\def\e{{\eta}}

\begin{document}

\leftline{\it Kiss Lajos b\'acsi, a Fizika Tan\'ara eml\'ek\'ere
\hfill {\rm hep-ph/9511404}}
\leftline{\it Dedicated to the memory of L. Kiss, Teacher of Physics}
\rightline{\null}
\rightline{\null}
\centerline{\normalsize\bf BOSE-EINSTEIN CORRELATIONS FOR JETS AND}
\baselineskip=22pt
\centerline{\normalsize\bf HEAVY ION REACTIONS
%\footnote{ Contribution to the XXV-th International Symposium on Multiparticle
%Dynamics, Stara Lesna, Slovakia, Sept. 12-16 1995 }
 }

%\vfill
%\vspace*{0.6cm}
\centerline{\footnotesize T. CS\"ORG\H O}
\baselineskip=13pt
\centerline{\footnotesize\it Department of Physics, Columbia University,
	538 West 120th, New York, NY 10027}
\centerline{\footnotesize\it MTA KFKI RMKI, H -- 1525 Budapest 114,
POB 49, Hungary}
\baselineskip=12pt
\centerline{\footnotesize E-mail: csorgo@sgiserv.rmki.kfki.hu}
\vspace*{0.3cm}
\centerline{\footnotesize and}
\vspace*{0.3cm}
\centerline{\footnotesize B. L\"ORSTAD}
\baselineskip=13pt
\centerline{\footnotesize\it Physics Institute, University of Lund,
Professorsgatan 1, S -- 223 63 Lund, Sweden}
\baselineskip=12pt
\centerline{\footnotesize E-mail: bengt@quark.lu.se}

%\vfill
\vspace*{0.9cm}
\abstracts{
Bose-Einstein correlations and invariant momentum distributions
are analyzed for longitudinally expanding finite systems,
like jets in elementary particle collisions or
systems created in high energy heavy ion reactions.
}

%\vspace*{0.6cm}
\normalsize\baselineskip=15pt
\setcounter{footnote}{0}
\renewcommand{\thefootnote}{\alph{footnote}}
\section{Introduction}
Bose-Einstein correlations are in general not measuring the whole
geometrical sizes of big and expanding finite systems\cite{nr,1d,3d}
since the expansion may result in strong correlations between space-time
and momentum space variables not only in the longitudinal,
but in the transverse and temporal
directions, too\cite{3d}.

Where have all the geometrical sizes gone?
One can show\cite{1d,3d}, that
they are disguised in the invariant momentum distribution
of the bosons in case they cancel from the
radius parameters of the Bose-Einstein correlation function
(BECF).

We shall briefly review herewith the results presented in
refs.\cite{nr,1d,3d,halo} which shall be appended with
an application to jet size determination.

\section{Wigner Function Formalism }
The two-particle inclusive correlation function is defined and
approximately expressed in the Wigner function formalism as
\ben
C(\Delta k;K) & = & {\dst \langle n(n - 1) \rangle \ov \langle n \rangle^2}
                {\dst N_2 ({\bf p}_1,{\bf p}_2)
                \ov N_1({\bf p}_1) \, N_1({\bf p}_2) }
                 \simeq
                1 + {\displaystyle\strut \mid \tilde S(\Delta k , K) \mid^2
                                \ov
                        \mid \tilde S(0,K)\mid^2 }.
\enn
In the above line,
the Wigner-function formalism\cite{pratt_csorgo,zajc,chapman_heinz}
 is utilized assuming fully chaotic (thermalized) particle emission.
The covariant Wigner-transform of the source density matrix,
 $S(x,p)$ is a quantum-mechanical  analogue of the classical
 probability that a boson is produced at a given
$ x^{\mu} = (t, {\bf r}) = (t,r_x,r_y,r_z)$
with $p^{\mu} = (E, {\bf p}) = (E, p_x, p_y, p_z)$.
The auxiliary quantity
$ \tilde S(\Delta k , K )  =  \int d^4 x S(x,K) \exp(i \Delta k \cdot x )$
appears in the definition of the BECF,
with $\Delta k  = p_1 - p_2$  and $K  = {(p_1 + p_2) / 2}$.
The single-  and two-particle inclusive momentum distributions (IMD-s)
 are given by
\ben
	N_1({\bf p})  =
	{\dst E \over \sigma_{tot}} {\dst d \sigma \ov d{\bf p}}
		= \tilde S(\Delta k = 0, p),
	\quad & {\mbox{\rm and}} & \quad
	N_2({\bf p_1},{\bf p_2})  =  {\dst E_1 E_2 \over \sigma_{tot}}
                 {\dst d \sigma \ov d{\bf p}_1 \, d{\bf p}_2},
\enn
where $\sigma_{tot}$ is the total inelastic cross-section.
Note that in this work we utilize the following
normalization of the emission function\cite{halo}:
$\int {\dst d^3{\bf p}\ov E}  d^4x S(x,p) = \langle n \rangle $.

\section{Effects from Large Halo of Long-Lived Resonances}
If the bosons originate from a core which is surrounded by
a halo of long-lived resonances, the IMD and the BECF can be
calculated in a straightforward manner. The detailed
description is given in ref.\cite{halo},
here we review only the basic idea.

If the emission function can be approximately divided into
two parts, representing the core and the halo,
        $ S(x;K)  =  S_{c}(x;K) + S_{h}(x;K) $
and if the halo is characterized
by large length-scales so that     $\tilde S_h(Q_{min};K) $
$ << \tilde S_c(Q_{min};K)$
at a finite experimental resolution of $Q_{min} \ge 10 $ MeV,
then the IMD and the BECF reads as
\ben
        N_1({\bf p}) & = &  N_{1,c}({\bf p}) +  N_{1,h}({\bf p}), \\
        C(\Delta k;K) & = & 1 + \lambda_*
              {\dst \mid \tilde S_{c}(\Delta k, K) \mid^2 \ov
                        \mid \tilde S_{c}(0,K) \mid^2 },
			\label{e:chalo}
\enn
where $N_{1,i}({\bf p})$ stands for the IMD
of the halo or core for $i = h,c$ and
\ben
        \lambda_* & = & \lambda_*(K = p) = \left[{\dst N_{1,c}({\bf p}) \ov
                                 N_{1}({\bf p})} \right]^2.
\enn
Thus within the core/halo picture the phenomenological $\lambda_*$
parameter can be obtained in a natural manner at a given finite
resolution of the relative momentum.
This parameter has been
introduced to the literature by Deutschmann long time ago\cite{deutschmann}.
In the core/halo picture, the effective
or measured intercept parameter $\lambda({\bf p})$ {\it can be interpreted}
as the {\it momentum dependent} square of the ratio of the IMD
of the core to the IMD of all particles emitted.

\section{General Considerations and Results}
We are considering jets in elementary particle
reactions or high energy heavy ion reactions,
which correspond to systems undergoing an approximately
boost-invariant longitudinal expansion.
For fully boost-invariant longitudinal expansions,
the emission function may depend on
boost-invariant variables only. These are
defined as $\tau= \sqrt{t^2 - r_z^2}, \, $
 $\eta = 0.5 \ln[(t+z)\,/\,( t-z)],\, $
$m_t = \sqrt{E^2 - p_z^2},\, $
$y = 0.5 \ln[(E + p_z )\,/\,(E - p_z)]\,$
and $r_t = \sqrt{r_x^2 + r_y^2}$.
For finite systems,
 the emission function may depend on $\eta - y_0$
too, where $y_0$ stands for the mid-rapidity.
Approximate boost-invariance is recovered in the
$\mid \eta - y_0 \mid << \Delta y$ region, where
the width of the rapidity distribution is denoted by $\Delta y$.
In terms of these variables the emission function can be rewritten
as
\ben
        S_c(x;K) \, d^4x & = & S_{c,*}(\tau,\eta,r_x,r_y) \,
                                d\tau \, \tau_0 d\eta \, dr_x \, dr_y.
\enn
The subscript $_{*}$ indicates that the functional form
of the source function is changed, and it stands for a
dependence on $K$ and $y_0$ also.

In the standard HBT coordinate system\cite{bertsch},
the mean and the relative momenta are
$K = (K_0,K_{\o},0,K_L)$ and $ \Delta k = (Q_0,Q_{\o},Q_{side},Q_L)$.
Note that the $_{side}$ component of the mean momentum
vanishes by definition\cite{bertsch,3d}.
Since the particles are on mass-shell, we have
$0  =  K \cdot \Delta k = K_0 Q_0 - K_L Q_L - K_{\o} Q_{\o}.$
\begin{figure}
\vspace*{13pt}
          \begin{center}
          \leavevmode\epsfysize=3.0in
          \epsfbox{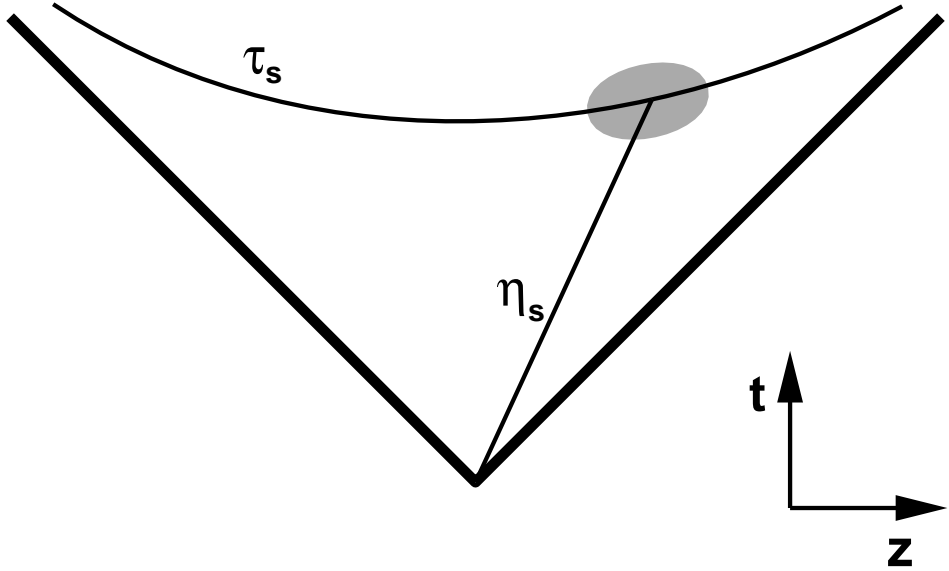}
          \end{center}
\caption{Emission of particles with a given momentum
is centered around $\tau_s$ and $\eta_s$
for systems undergoing boost-invariant longitudinal expansion,
as indicated by the shaded area.  }
\label{fig:1}
\end{figure}
Introducing $\beta_L  = K_L / K_0$ and $\beta_\o = K_\o / K_0$,
the energy difference $Q_0$ can thus be expressed as
\ben
        Q_0 & = & \beta_L Q_L + \beta_{\o} Q_\o. \label{e:11}
\enn
If the emission function has a such a structure that
it is concentrated in a narrow region around
$(\tau_s,\eta_s)$ in the $(\tau,\eta)$ plane,
then one can evaluate the BECF in terms of variables
$\tau$ and $\eta$ by utilizing the expansion
\ben
        \Delta k \cdot x & = & Q_0 t - Q_\o r_x - Q_\s r_y - Q_L r_z \simeq \\
		 \null & \null &
                Q_\t \t - Q_\o r_x - Q_\s r_y - Q_\e \t_s (\eta - \eta_s).
\enn
The coefficients of the $\t$ and the $\t_s (\e - \e_s)$ are new variables
given by
\ben
        Q_\t & = & Q_0 \cosh[\eta_s] - Q_L \sinh[\eta_s] =
	 % \\ \null & \null &
        (\beta_t Q_\o + \beta_L Q_L)  \cosh[\eta_s] - Q_L \sinh[\eta_s], \\
        Q_\e & = &  Q_L \cosh[\eta_s] - Q_0 \sinh[\eta_s] =
	 % \\ \null & \null &
        Q_L \cosh[\eta_s] - (\beta_t Q_\o + \beta_L Q_L)  \sinh[\eta_s].
\enn
In terms of these new variables the BECF reads as
\ben
        C(\Delta k;K) \simeq
                1 + {\displaystyle\strut \mid \tilde S(\Delta k , K) \mid^2
                                \ov
                        \mid \tilde S(0,K)\mid^2 }
                \simeq 		1 + \lambda_*(K)
                {\dst \mid \tilde S_{c,*}
				(Q_\t, Q_\e, Q_\o, Q_\s) \mid^2 \ov
                        \mid \tilde S_{c,*}(0,0,0,0) \mid^2}.
		\label{e:cgen}
\enn
At this level, the shape of the BECF can be rather complicated,
it may have non-Gaussian, non-factorizable structure.
Gaussian approximation to eq.~(\ref{e:cgen}) may break down
as discussed in more detail in the  Appendix of ref.\cite{3d}.

\section{Mixing Angle for HBT}
In the experimental analysis, one of the most frequently\cite{bengt}
but not exclusively\cite{ua1} applied parameterization of the
BECF is some version of a Gaussian approximation. The out-longitudinal
cross-term of BECF has also been discovered in this
context recently\cite{uli_sh}.
In order to identify how this term may come about, let us assume that
\ben
        S_{c,*}(\tau,\eta,r_x,r_y) & = & H_*(\t) \, G_*(\e) \, I_*(r_x,r_y).
        \label{e:fact}
\enn
In Gaussian approximation one also assumes that
\ben
H_*(\tau ) & \propto & \exp(-(\t - \t_s)^2 / (2 \Delta \t_*^2) \, ),
        \label{e:hst} \\
G_*(\eta ) & \propto & \exp(-(\eta - \eta_s)^2 / (2 \Delta\e_*^2) \, ),
        \label{e:gst} \\
I_*(r_x,r_y) & \propto & \exp(-(\, (r_x - r_{x,s})^2 + (r_y - r_{y,s})^2 )
                        / (2 R_*^2) \, ).
\label{e:ist}
\enn
The corresponding BECF is given by a diagonal form as
\ben
C(\Delta k;K) = 1 + \lambda_* \,
                \exp( - Q_\t^2 \Delta\tau_*^2 - Q_\e^2 \t_s^2 \Delta\e_*^2 -
                 Q_t^2 R_*^2). \label{e:c2di}
\enn
This diagonal form shall be transformed to an off-diagonal
one if one introduces the kinematic relations between
the variables $ Q_\t,  Q_\e$ and the variables $Q_\o,Q_L$.
In the HBT coordinate system\cite{bertsch} one finds
\ben
        C(\Delta k;K) & = & 1 + \lambda_* \exp( -R_\s^2 Q_\s^2
			-R_\o^2 Q_\o^2
                  -R_L^2 Q_L^2  -2 R^2_{\o,L} Q_\o Q_L),
	% \\
	% \times \\
        % \null & \null &
        % \times \exp( - 2 R^2_{\o,L} Q_\o Q_L )
	\label{e:crbecf} \\
        R_\s^2 & = & R_*^2, \label{e:rs} \\
        R_\o^2 & = &  R_*^2 + \delta R_\o^2, \label{e:ro} \\
        \delta R_\o^2 & = & \beta_t^2 ( \cosh^2[\e_s] \Delta\t_*^2 +
                                        \sinh^2[\e_s] \t_s^2 \Delta\e_*^2),
                                \label{e:dro}\\
        R_L^2 & = &
        (\beta_L \sinh[\eta_s] - \cosh[\e_s])^2 \t_s^2 \Delta\e_*^2  +
	% \\ \null & \null &
                        ( \beta_L \cosh[\eta_s] - \sinh[\e_s])^2 \Delta\t_*^2,
                        \label{e:rl}\\
        R_{\o,L}^2 & = & (\beta_t \cosh[\eta_s]
        ( \beta_L \cosh[\eta_s] - \sinh[\e_s])) \Delta\t_*^2 + \nonumber\\
                \null & \null &                 (\beta_t \sinh[\eta_s]
                (\beta_L \sinh[\eta_s] - \cosh[\e_s]) ) \t_s^2 \Delta\e_*^2.
                \label{e:rol}
\enn
Note that the effective temporal duration, $\Delta\t_*$
and the effective longitudinal size, $\tau_s\Delta\eta_*$
appear in a mixed form in the BECF parameters $\delta R_\o^2$,
$R_L^2$ and $R_{\o,L}^2$, and their mixing is controlled by
the value of the parameter $\eta_s$.
These results simplify a lot\cite{3d} in the LCMS,
the Longitudinally Co-Moving System\cite{LCMS},
where $\beta_L = 0$:
\ben
        \delta R_\o^2 & = & \beta_t^2 ( \cosh^2[\e_s] \Delta\t_*^2 +
                        \sinh^2[\e_s] \t_s^2 \Delta\e_*^2), \label{e:ldro} \\
        R_L^2 & = & \cosh^2[\e_s] \t_s^2 \Delta\e_*^2 +
                        \sinh^2[\e_s] \Delta\t_*^2, \label{e:lrl} \\
         R_{\o,L}^2 & = & - \beta_t \sinh[\e_s] \cosh[\eta_s] (  \Delta\t_*^2 +
                         \t_s^2 \Delta\e_*^2). \label{e:lrol}
\enn
Let us define the Longitudinal Saddle-Point System (LSPS)
to be the frame where $\eta_s(m_t) = 0$. In LSPS one finds that
\ben
       \delta R_\o^2 & = & \beta_t^2 \Delta\t_*^2, \\
         R_L^2 & = & \t_s^2 \Delta\e_*^2 + \beta_L^2 \Delta\tau^2_*, \\
         R_{\o,L}^2 & = & \beta_t \beta_L \Delta\tau_*^2.
	\label{e:lspsrl}
\enn
Introducing $Q_0 = \beta_t Q_\o + \beta_L Q_L$
and $Q_t = \sqrt{ Q_\o^2 + Q_\s^2}$ the BECF can be rewritten in LSPS as
\ben
C(\Delta k;K) & = & 1 + \lambda_*
                \exp( - \Delta\tau^2_* Q_0^2 - \t_s^2 \Delta\e_*^2 Q_L^2 -
                R_*^2 Q_t^2).
\enn
Thus the out-long cross-term can be diagonalized in the LSPS
frame\cite{nix,3d}. The cross term should be small in LCMS
if $\eta_s^{LCMS} << 1$ i.e. if $\mid y - y_0 \mid << \Delta y$~\cite{3d}.
Since the size of the cross-term is controlled by the value
of $\eta_s$ in any given frame, it follows that
 $\eta_s$ is the {\it cross-term generating
hyperbolic mixing angle}\cite{3d}
for cylindrically symmetric, longitudinally expanding finite
systems which satisfy the
factorization of eq.~(\ref{e:fact}).

\section{Decaying Lund Jets}
An ultra-relativistic jet corresponds to a boost-invariant
longitudinally expanding finite system, with  strong
correlation among longitudinal space-time and momentum-space
variables. Finite correlation lengths are created by the
random fluctuation of the break-up points of the hadronic
string. Local left-right symmetry of the jet fragmentation
prescribes the following proper-time distribution\cite{JET74}:
\ben
H(\tau) d\tau = {\dst 2 \ov {\rm \Gamma} (1 + a)}
        b^{1+a} (\kappa\tau)^{2a+1} \exp(-b(\kappa\tau)^2) \kappa
	d\tau
\enn
where $\kappa \approx 1$ GeV/fm,
$a = 0.3$ and $b = 0.58$ GeV$^{-2}$ are the default parameters
of JETSET7.4~\cite{JET74}.
In Gaussian approximation to the BECF one obtains
\ben
\tau_s  =  \langle \tau \rangle = {\dst 1 \ov \sqrt{ b \kappa^2} } \,
		{{\rm \Gamma} (3/2 + a) \ov {\rm \Gamma} (1 + a) },
	% \\
	\quad & \mbox{\rm and} & \quad
\Delta\tau_*^2  =  {\dst 1 \ov b \kappa^2}
		\l( 1 + a - { {\rm \Gamma}^2 (3/2 + a) \ov
                {\rm \Gamma}^2 (1 + a) } \r).
\enn
The BECF for prompt pions may become measurable
as discussed in ref.\cite{Verbeure}.
If the prompt pions cannot be separated, the above
expressions need to be corrected for the
resonance decay effects which may increase both
the mean and the variance of the distribution.
The transverse radius parameter for direct pions
is $   R_{side}  =  R_*  \approx R_{string}$
and the correlation length between space-time rapidity
and momentum space rapidity is to be evaluated
numerically.
This can be performed by utilizing the
Lund mapping of momentum space to space-time
along the lines of refs.\cite{LCMS,JET74}.
When resonance decays were switched off,
one obtained\cite{LCMS} $\Delta\eta_* \simeq 0.5$
using JETSET6.3 default parameters of
$a = 0.5 $ and $b = 0.9$ GeV$^{-2}$.
This yields $ R_L \approx \tau_0 \Delta\eta_* \simeq 0.6 $ fm
and $ \Delta\tau_* \approx 0.5 $ fm/c for prompt pions.
When resonance decays are switched on,
the width of the $G_*(\eta-y)$ distribution (for all pions)
increased to $\Delta\eta_* \approx 1$ unit.
Should the measured BECF for direct pions be fitted with
the expression
\ben
C(\Delta k,K) \simeq 1 + \lambda_* \exp( - Q_{\tau}^2 \Delta\tau^2_* -
Q_L^2 R_L^2 - Q_t^2 R_{string}^2 ),
\enn
the calculated numbers could be contrasted to data
and some information on the transverse size of the
string may also become available.

It should be clear that even a successful description
of the BECF for particles from a decaying string,
like the famous Andersson-Hofmann model\cite{bo}
does not directly reveal the total longitudinal
size of the string because the BECF is dominated by
the correlation length between space-time and momentum
space rapidity\cite{1d}.

In ref.\cite{1d} one of us argued that the total longitudinal
sizes of expanding finite systems may become measurable
with the help of a combined analysis of IMD and BECF measurements.
According to ref.\cite{1d}, the space-time
rapidity distribution of the boson source can be measured
as the asymptotic large transverse mass limit of the
invariant momentum distribution.
%Indeed, both an iterative
%jet-fragmentation mechanism and a locally thermalized
%momentum distribution describing a finite system with scaling
%longitudinal flow profile predicts that the correlations between
%the variables $(y ,\eta)$
%shall become proportional to $\delta(y -\eta)$ in the large
%transverse mass limit.

These results may be considered as
first steps into a new direction of jet size determination
with the help of combined IMD and BECF measurements.

\section{Geometrical vs. Thermal Length Scales for Heavy Ions}
For high energy heavy ion reactions, we model the emission function
of the core with an emission function described in detail
in ref.\cite{3d}. This corresponds to
        a Boltzmann approximation to the local momentum distribution
        of a longitudinally expanding finite system which expands
        into the transverse directions with a transverse flow,
        which is non-relativistic at the maximum of particle emission.
The decrease of the temperature distribution $T(x)$
in the transverse direction is
controlled by a parameter $a$,
the strength of the transverse flow is controlled by a parameter $b$.
        A parameter $d$ controls the strength of the change of the
        local temperature during the course of particle emission\cite{3d}.
If all these parameters vanish, $a = b = d = 0$, one recovers the case
of longitudinally expanding finite systems with $T(x) = T_0$
with no transverse flow, as discussed in ref.\cite{1d},
if $a = d = 0 \ne b$ the model of ref.\cite{uli_sh} is obtained.

        The parameters of the correlation function are related
	by eqs.~(\ref{e:crbecf}-\ref{e:lspsrl}) to
	the parameters $R_*$, $\tau_*$ and $\tau_s \Delta\eta_*$
	which in turn are given by
\ben
{\dst 1 \ov R_*^2 } & = &
                {\dst 1 \ov R_G^2} +
                {\dst 1 \ov R_T^2 } \cosh[\eta_s],  \\
{\dst 1 \ov \Delta \eta_*^2} & = &  {\dst 1 \ov \Delta \eta^2 } +
                {\dst 1 \ov \Delta \eta_T^2} \cosh[\eta_s] -
                        {\dst 1 \ov \cosh^2[\eta_s]}, \\
{\dst 1 \ov \Delta \tau_*^2 } & = &
                {\dst 1 \ov \Delta\tau^2} + {\dst 1 \ov \Delta\tau_T^2}
                        \cosh[\eta_s].
\enn
Here the geometrical sizes are given by $R_G$, the transverse size,
$\Delta \eta$, the width of space-time rapidity distribution and
$\Delta \tau$, the duration around the mean emission
time $\tau_s = \tau_0$.  The hyperbolic mixing angle $\eta_s\approx 0$
at midrapidity $y_0$~\cite{3d}, where also the
out-long cross-term\cite{uli_sh} vanishes.
The thermal length-scales (subscript $_T$) are given by
\ben
 R_T^2  =  { \displaystyle\strut \tau_0^2 \over
        \displaystyle\strut a^2 + b^2 }
         { \displaystyle\strut T_0 \over \displaystyle\strut M_t},
	\quad &
 \Delta\eta_T^2  =  {\dst T_0 \ov M_t},
	& \quad
 \Delta\tau_T^2  =  {\dst \tau_0^2 \ov d^2} {\dst T_0 \ov M_t}.
\enn
The transverse mass of the pair is denoted by $M_t = \sqrt{K_0^2 - K_L^2}$.

These analytic expressions indicate that the BECF views only
a part of the space-time volume of the expanding systems,
which implies that even a complete measurement of the
parameters of the BECF as a function of the mean momentum $K$
may not be sufficient to determine uniquely
 the underlying phase-space distribution.

\begin{figure}
%\vspace{-0.5truein}
          \begin{center}
          \leavevmode\epsfysize=3.0in
          \epsfbox{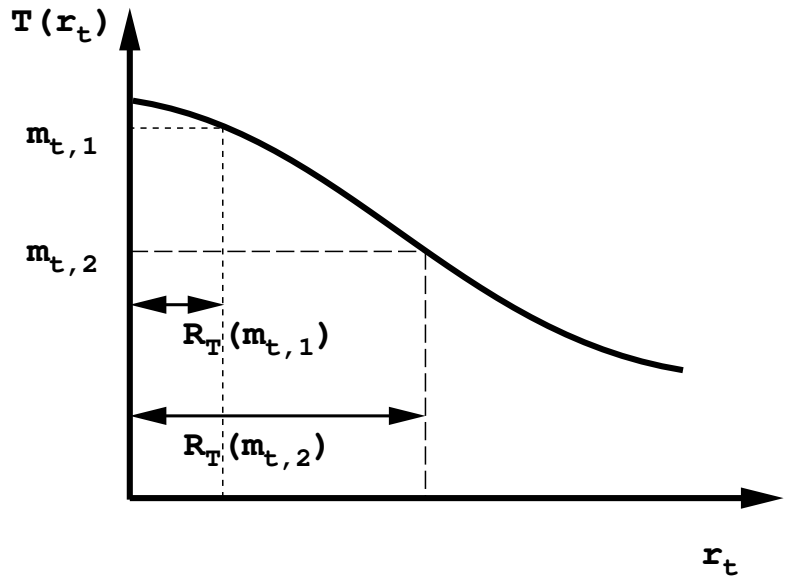}
          \end{center}
\caption{
The temperature gradients in the transverse
direction create a transverse mass dependent  effective
thermal radius parameter. In this illustration,
the effective size of the region where particles with a given $m_t$
are emitted from is decreasing
with increasing values of $m_t$. Note that the transverse
flow gradients may result in a similar effect$^{3,7}$,
 not indicated on this illustration.
}
\label{fig:2}
\end{figure}

It is timely to emphasize at this point that the parameters
of the Bose-Einstein correlation function coincide with the
(rapidity and transverse mass dependent)
{\it lengths of homogeneity}\cite{sinyukov}
in the source, which can be identified with that region
in coordinate space where particles with a given momentum are emitted
from. The lengths of homogeneity
for thermal models can be obtained from basically two type of scales
referred to as 'thermal' and 'geometrical' scales.

The thermal scales originate from the factor
$\exp(- p \cdot u(x)  /T(x))$,  where $u(x)$ is the four-velocity field.
Figure 2 and 3 illustrate how the temperature
changes in the transverse or temporal directions induce transverse mass
dependent thermal radius or thermal duration parameters.
This is to be contrasted to the 'geometrical'
scales, which originate from the $\exp( \mu(x) /T(x) )$ factor which
controls the density distribution\cite{3d}. Here $\mu(x)$ stands
for the chemical potential.

\begin{figure}
%\vspace*{13pt}
          \begin{center}
          \leavevmode\epsfysize=3.0in
          \epsfbox{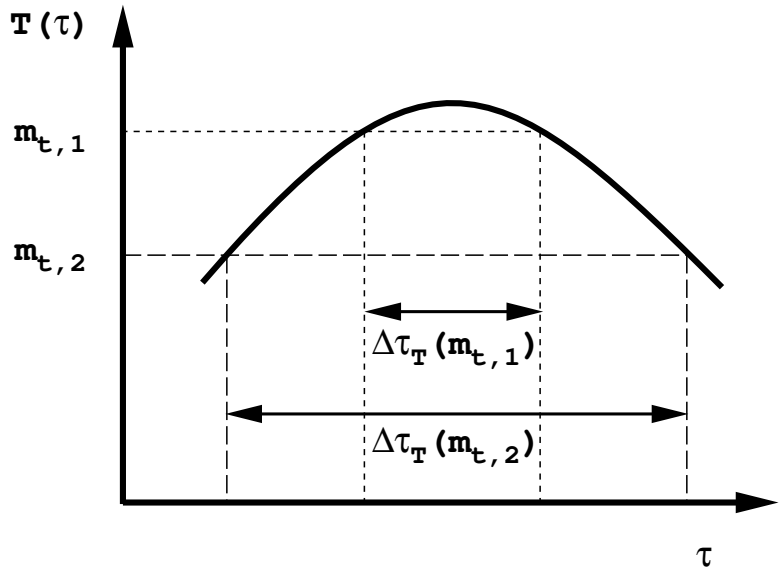}
          \end{center}
\caption{
Temporal changes of the local temperature
create a transverse mass dependent effective thermal duration,
which is decreasing
with increasing values of $m_t$ in this illustration.
}
\label{fig:3}
\end{figure}

\section{Dynamically Generated Vanishing Life-Time Parameter}
As a consequence of the possibility for temporal changes
of the local temperature, we find that the effective duration
of the particle emission $\Delta\tau_*$ becomes
transverse mass dependent and for sufficiently large
values of the transverse mass this parameter may become vanishingly small.
The reason for this new effect is rather simple:
Particles with a higher transverse mass are effectively emitted
in a time interval when the local temperature (boosted by the transverse
flow) is higher than the considered value for $m_t$.
If the local temperature changes during the course of particle
emission, the effective emission time for high transverse mass
particles shall be smaller than the effective emission time of
particles with lower transverse mass values.

For a more detailed analysis
of the model the interested reader is referred to
ref.\cite{3d,japan}, where it is pointed out that under certain
conditions the parameters of the Bose-Einstein correlation
function may obey an $M_t$-scaling: $R_\s \simeq R_\o \simeq R_L
\propto 1/\sqrt{M_t}$.

\medskip
\leftline{\bf Acknowledgments}
Cs. T. would like to thank the Organizers
for invitation and local support, M.Gyulassy and B. L\"orstad
for kind hospitality at Columbia and Lund University.
This work has been supported by the HNSF grants
OTKA - F4019 and W01015107.

\end{document}